\documentclass[manuscript]{aastex}

\shorttitle{Zero metal massive stars}
\shortauthors{Limongi and Chieffi}

\begin{document}

\title{Presupernova evolution and explosive nucleosynthesis of zero metal massive stars}

\author{M. Limongi\altaffilmark{1,3,4}}
\email{marco.limongi@oa-roma.inaf.it}

\and

\author{A. Chieffi\altaffilmark{2,4}}
\email{alessandro.chieffi@inaf.it}

\altaffiltext{1}{Istituto Nazionale di Astrofisica - Osservatorio Astronomico di Roma, Via Frascati 33, I-00040, Monteporzio Catone, Roma, Italy}
\altaffiltext{2}{Istituto Nazionale di Astrofisica - Istituto di Astrofisica Spaziale e Fisica Cosmica,
              Via Fosso del Cavaliere, I-00133, Roma, Italy}              
\altaffiltext{3}{Kavli Institute for the Physics and Mathematics of the Universe, Todai Institutes for Advanced Study, the University of Tokyo, Kashiwa, Japan 277-8583 (Kavli IPMU, WPI)}              
\altaffiltext{4}{Centre for Stellar \& Planetary Astrophysics, School of Mathematical Sciences, P.O. Box, 28M, Monash University, Victoria 3800, Australia}

\begin{abstract}
We present a new set of zero metallicity models in the range 13-80 $\rm M_\odot$ together to the associated explosive nucleosynthesis. These models are fully homogeneous with the solar metallicity set we published in Limongi \& Chieffi (2006) and will be freely available at the web site http://www.iasf-roma.inaf.it./orfeo/public{\_}html.
A comparison between these yields and an average star that represents the average behavior of most of the very metal poor stars in the range $\rm -5.0<[Fe/H]<-2.5$ confirms previous findings that only a fraction of the elemental [X/Fe] may be fitted by the ejecta of $\it standard$ core collapse supernovae.
\end{abstract}

\keywords{early universe --- Galaxy: abundances --- nuclear reactions, nucleosynthesis, abundances --- stars: abundances --- stars: evolution --- supernovae: general}

\section{Introduction} \label{intro}
The current wisdom that the iron content in stars increases with the cosmic age
places the birth of the extremely iron poor stars at the very early epochs of the evolution of the
Universe, at redshifts larger than 5 \citep{cb03}. The most iron poor stars likely formed
from gas clouds enriched by very few stellar generations or even from just the first generation of stars \citep{ssa07}. 
Then, a detailed analysis of the surface chemical
composition of these objects, which should reflect the composition of the gas out of which these
stars formed,  represents a powerful tool to investigate the nature of the first stellar generation(s).
 
Detailed chemical abundance measurements, obtained over the years by several groups 
(HK survey, Beers, Preston \& Shectman 1992; Hamburg-ESO (HES) survey, Christlieb 2003, 
Christlieb et al. 2008; Sloan Extension for Galactic Understanding and Exploration (SEGUE), Yanny et al. 2009) led to the following main results. (1) The metallicity distribution function (MDF) is characterized by a rapid decrease of the number
of stars with decreasing metallicity, by a sharp cut off at $\rm [Fe/H]\simeq -4$ (we adopt the standard notation $\rm [Fe/H]=log_{10}(Fe/H)-log_{10}(Fe/H)_\odot$), and by the presence of only two objects at $\rm [Fe/H]\sim -5.3$, i.e. about a factor of 30 below the sharp cut off at $\rm [Fe/H]\simeq -4$. These results led to some speculations about
a possible metallicity "gap" between $\rm [Fe/H]\simeq -4$ and $\rm [Fe/H]\simeq -5.3$, "gap" which could have profound implications
on both the nature of the first stars and the mechanism of enrichment of the pristine material.
(2) The existence of a well defined abundance pattern shared by the majority of the stars in the
metallicity range $\rm -4.0<[Fe/H]<-3.0$ \citep{cayreletal04}. (3) The existence of a fraction
of stars showing extremely high overabundances of C relative to Fe (in most cases associated to large overabundances of N and O) 
that increases as the metallicity decreases \citep{bc05}. This tendency, coupled to the fact that the two most
metal poor stars known to date have a $\rm [Fe/H]\sim -5.3$ (HE0107-5240 Chriestlib et al. 2002, Christlieb et al. 2004; HE1327-2326: Frebel et al. 2005, Aoki et al. 2006) and show an enormous overabundance of C, N and O 
relative to Fe, suggesting that this feature could be ubiquitous at these low metallicities.
(4) The existence of a number of subclasses among the "normal" and the C-rich stars on the basis of the
enhancement of r-process and/or s-process elements \citep[see][]{bc05}.

The quite recent discovery of HE0557-4840 \citep{norrisetal07}, a C-rich star with [Fe/H]=-4.75, 
questioned the existence of the metallicity gap between [Fe/H]=-4 and [Fe/H]=-5.3 but reinforced the
idea that, below a given metallicity, all stars are strongly enriched in C relative to Fe. 

\begin{figure}
\epsscale{.80}
\plotone{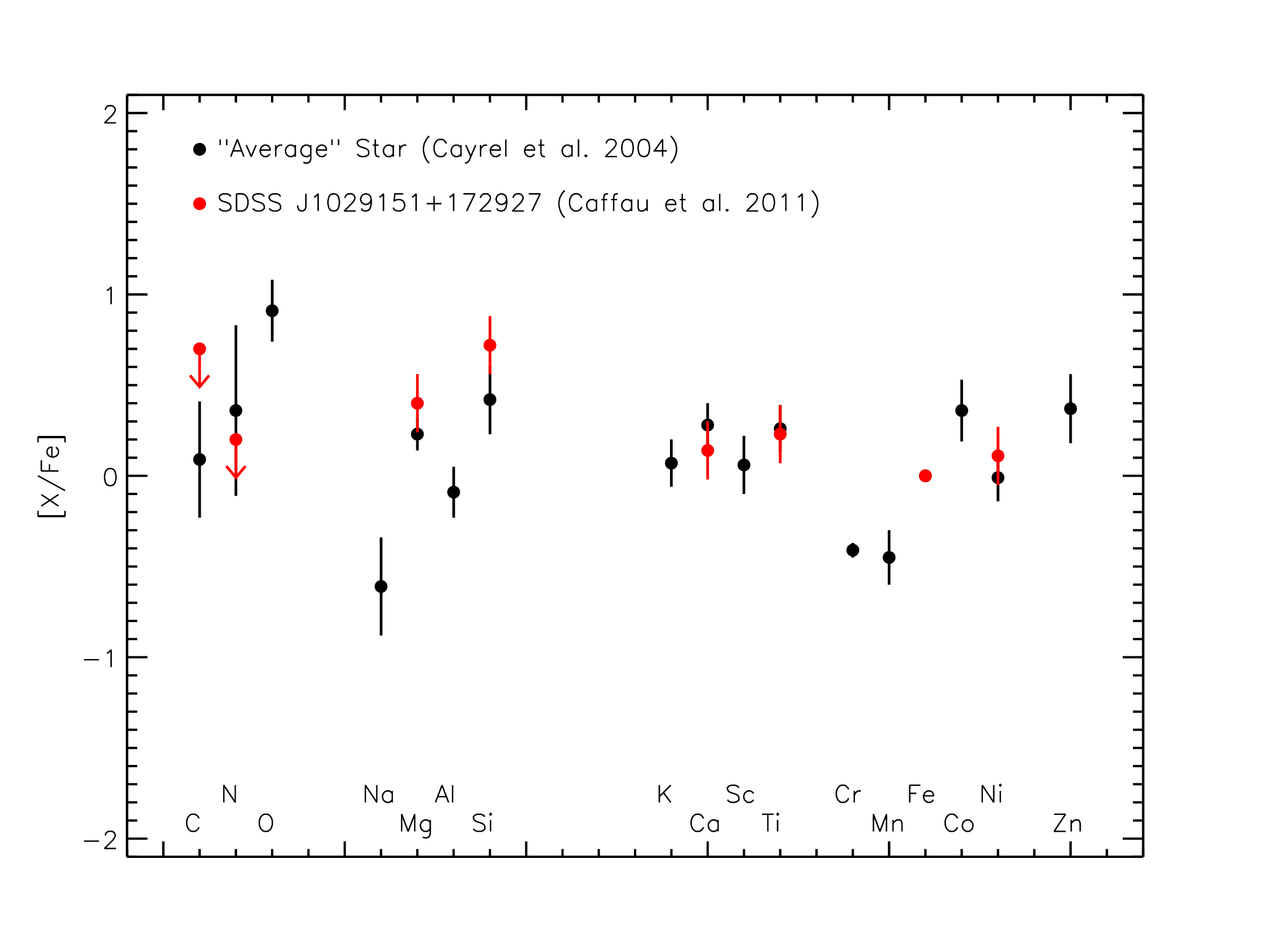}
\caption{Comparison between the abundance patterns of the average star \citep{cayreletal04,spiteetal05} and of the 
SDSS J1029151+172927 star \citep{caffauetal11}}.\label{compstars}
\end{figure}

Very recently \citet{caffauetal11} discovered the star SDSS J1029151+172927 ([Fe/H]=-4.99).
This star shows a normal abundance pattern, i.e., similar to that shared by the majority of the stars in the metallicity range $\rm -4.0<[Fe/H]<-3.0$, and a normal [C/Fe] ratio (Figure \ref{compstars}).
This discovery calls into serious doubts that the strong overabundance of C with respect to Fe is ubiquitous at extremely low metallicities
and reinforces the idea that the common pattern shared by the majority of stars with  $\rm [Fe/H]\ge -4$ may
extend to much lower metallicities.

From the theoretical point of view, there have been many attempts in the last 10 years to interpret
the observed abundance pattern in the most iron poor stars in terms of one or more zero metallicity core collapse
supernovae \citep{lc02,cl02,lcb03,un02,un03,un05}. 
However, the discovery of many C-rich stars in the population of the extremely Fe poor stars led many authors
to consider the extremely metal poor asymptotic giant branch stars (AGB) and/or binary mass transfer \citep[][and references therein]{sudaetal11} as a possible cause of the large overabundances of C and N in these objects.

In the framework of the pure massive star scenario, despite the great efforts in modelling the presupernova evolution
as well as the explosive nucleosynthesis, at present no classical model is able to reproduce the overall observed abundance pattern of both the normal and the C-rich stars. In particular, although a single or a generation of zero metallicity massive stars may provide a good fit
to several elemental abundance ratios of the "average" star, i.e. of a representative star having an abundance
pattern common to the majority of the stars with $\rm [Fe/H]<-3.0$,
there are [X/Fe] log ratios for which the fit can be obtained only by means of specific "ad-hoc" assumptions.
For example, the large observed [Zn/Fe] ratio can be reproduced by energetic
explosions, with final kinetic energies up to ten times the typical ones ($10^{51}$ erg) \citep{un02};
the very low [Co/Fe] ratio obtained in the classical models can be significantly increased and reconciled with the observations 
by using energetic explosions $\it and$ an electron mole number in the complete explosive Si burning region artificially increased up to values larger than $\simeq 0.5$ \citep{un02}. By the way the [Co/Fe] ratio may be raised also by increasing the $\rm ^{12}C$ mass fraction at core He depletion by varying the mixing efficiency during central He burning and/or the rate of the $\rm ^{12}C(\alpha,\gamma)^{16}O$ cross section within the range of the currently accepted uncertainties \citep{cl02}.
The underabundances of Sc and Ti relative to Fe may be significantly enhanced by artificially reducing the density during the explosive burning, i.e. enhancing the $\alpha$-rich freeze out \citep{un05}. \cite{mn03} showed that an aspherical explosion can lead to a profound reduction of the density in the deep interior of the star but such a possibility must be investigated further. A similar situation is found when attempting to fit the observed abundance pattern of the C-rich most metal poor stars. The extremely high overabundance of the light elements relative to the iron peak ones could be explained in the framework of the explosion of a single supernova by assuming different combinations of explosion energy and efficiency of mixing and fallback \citep{un02} or in terms of an almost failed supernova, .i.e. a supernova experiencing a large fallback, exploding within an environment enriched by a generation of zero metallicity supernovae \citep{lcb03}. Very recently, also \citet{hw10} found that individual ultra metal poor stars could be fitted by a proper combination of explosion energy, mixing efficiency and IMF. Although there could be strong arguments in favour of one or another of these explanations, the need of different specific fine tunings of the various parameters to fit the abundance pattern of any single star may simply reflect the lack of fundamental physical processes in the presently available calculations or more simply that the basic assumption that the chemical composition observed on the surface of these stars comes just from one or more massive zero metallicity stars is wrong.

In this paper we present our latest zero metallicity massive star models in the range 13-80 $\rm M_\odot$ and their related explosive nucleosynthesis. These models differ from our previous set of zero metallicity stars published in \citet{cl04} because of a number of changes in both the stellar evolutionary code (numerical scheme, input physics, nuclear cross sections, nuclear network) and the explosive nucleosynthesis (see next section). These new versions of the hydrostatic code and of the hydrodynamical one have already been used to produce a new set of solar metallicity stellar models \citep{lc06} and hence the present models are the natural complement of those ones. We take advantage of these new models to compare our theoretical predictions with the latest observed abundances in the extremely metal poor stars to check whether the discrepancies between theory and observations, described above, still remain.


\section{Stellar evolution and explosive nucleosynthesis calculations} \label{codes}
This set of zero metallicity models extends in mass between 13 and 80 $\rm M_\odot$ and all models
are assumed to have a pristine Big Bang nucleosynthesis composition.
Their evolution has been followed from the pre-main sequence phase 
up to the onset of the iron core collapse by means of the latest, stable, version of the FRANEC stellar evolutionary code,
which is described in detail in \citet{lc06}.
The main improvements of this version with respect to the previous one, described in \citet{lc03} and adopted to
compute our previous set of zero metallicity massive star models, are the following.
First of all, the convective mixing and the nuclear burning are coupled together and solved simultaneously.
The mixing is described by a diffusion equation where the diffusion coefficient $D$ is 
given by $D=1/3 v_c l$, the convective velocity $v_c$ being computed in the framework of the mixing length theory. 
The nuclear cross sections have been updated with respect to those adopted in \citet{lc03} whenever possible. Table 1 
of \citet{lc06} shows the full reference matrix of all the processes taken into account in the network, together to its proper 
legend. The nuclear networks for the He and the advanced burning phases have been extended to 153 
and 282 elements (from H to $\rm ^{\rm 98}Mo$) respectively. In total 297 isotopes and about 3000 processes were 
explicitly included in the various nuclear burning stages. 
In the present calculations we have assumed 0.2 $H_p$ of overshooting at the top of the convective core during core H burning and the mass loss is switched off during all the evolutionary stages. Thus, these models and chemical yields are perfectly homogenous to the set of solar metallicity massive star models presented in \citet{lc06}. 

The hydro code adopted in this paper to compute the explosion of the mantle of the star and the associated explosive nucleosynthesis
has been significantly improved with respect to that adopted in \cite{lc03}.
More specifically, the explosion of the mantle of the star (i.e., all the zones above the iron core) is started by imparting instantaneously an initial 
velocity $v_0$ to a mass coordinate of $\rm \sim 1~M_\odot$ of the presupernova model, i.e., well within 
the iron core. The propagation of the shock wave that forms consequently is followed by means of an 
hydro code, fully developed by us, that solves the fully compressible reactive hydrodynamic equations using the piecewise parabolic 
method (PPM) of \citet{cw84} in the Lagrangean form. On the contrary, in \citet{lc03} we adopted a simpler 
forward time centered space scheme as described by \citet{rm67} and \citet{mb93}.
The initial velocity $v_0$ is properly tuned in order to obtain either a given final kinetic energy of the ejecta,
typically of the order of $\rm 10^{51}~erg$, or a specific mass cut or eject a given amount of $\rm ^{56}Ni$.
The chemical evolution of the matter is computed by coupling the same nuclear network 
adopted in the hydrostatic calculations to the system of hydrodynamic equations. 
The nuclear energy generation is neglected since we assume that it is 
always negligible compared to both the kinetic and the internal energies.

\section{Presupernova evolution}

The distinctive feature of the evolution of a massive zero metal star is the lack of CNO nuclei. This occurrence implies 
that the energy required to preserve the hydrostatic equilibrium cannot be provided by the CNO cycle. Since the PP chain cannot supply this energy at temperatures lower than $\sim10^{8}$ K or so, threshold temperature for the onset of the He burning \citep{lcs98}, the star is forced to reach such high temperature where a little bit of C is produced by a partial activation of the $3~\alpha$ nuclear process. The C produced by the $3~\alpha$ triggers the CNO cycle that can now provide enough energy to sustain the star. The net result is that the H burning partially overlaps, in temperature, to the He burning. While such an occurrence does not imply substantial differences in the main evolutionary properties of the star in core H burning \citep{lcs98}, it may largely affect the behavior of both the He and the H burning shells. More specifically, since the He and the H burning shells operate at almost the same temperature, the entropy barrier that develops at the He-H interface lowers considerably. As a consequence, a partial mixing between the He convective shell and the H rich envelope is not inhibited any more and it actually occurs not unfrequently. The protons ingested by the He convective shell give rise to a burst of nuclear energy, which in turn induces a rapid increase of the convective shell, and activate a sequence of reactions that lead to a primary production of N, Na and Mg \citep{lcb03}. Though the ingestion of protons in the active He burning region in these zero metallicity stars has been already widely found and discussed in the literature \citep{ww82,chieffietal01,fih90,hw10}, different results have been found by different authors mainly because of the complex interplay between the convective mixing and the nuclear burning (still very difficult to model properly).

\begin{figure}
\epsscale{.80}
\plotone{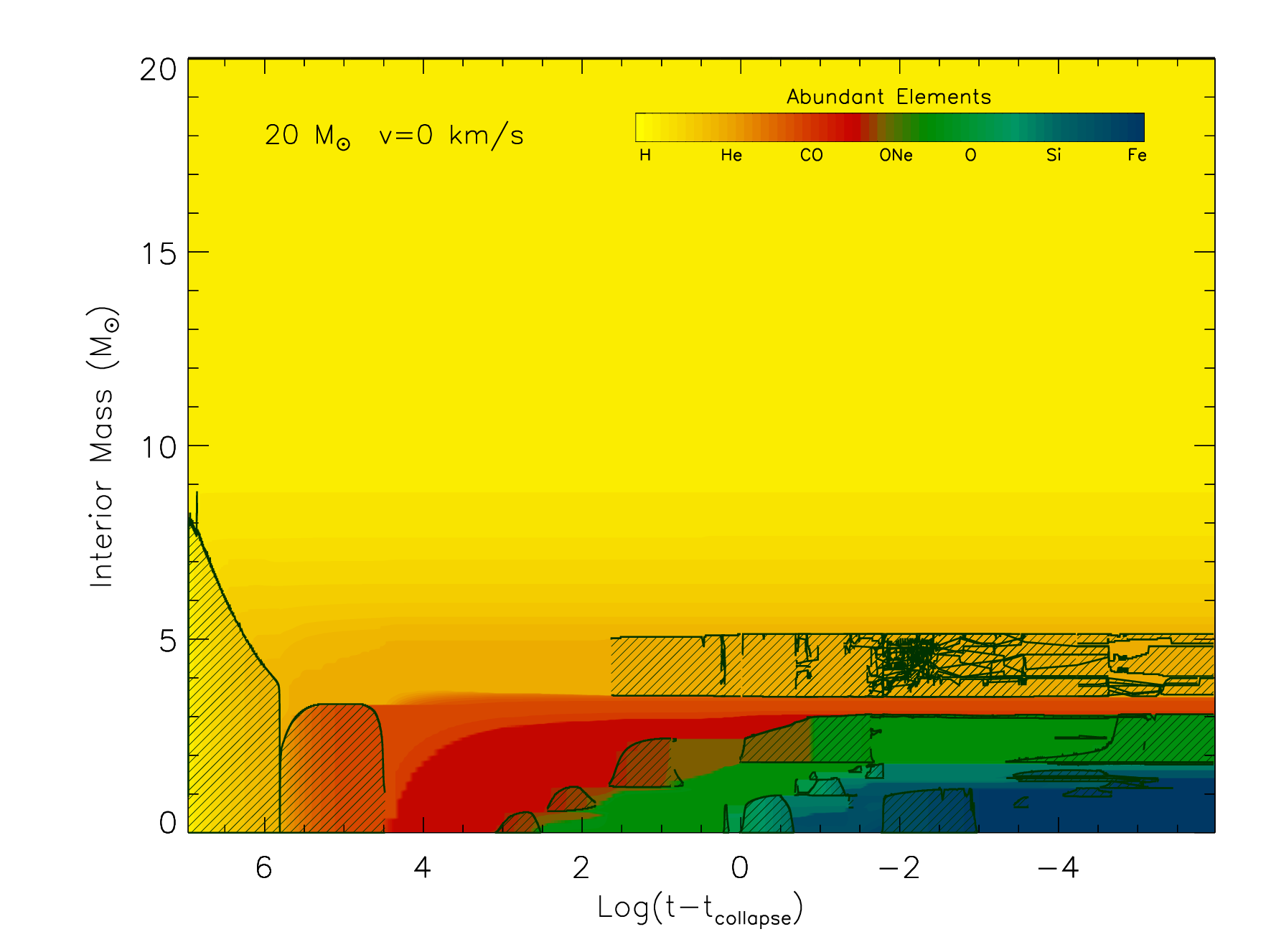}
\caption{Presupernova chemical and convective history of a $\rm 20÷M_\odot$. The shaded
regions correspond to the convective zones. The color gradient refers to the chemical composition as
reported in the colorbar}.\label{conv1}
\end{figure}

\begin{figure}
\epsscale{.80}
\plotone{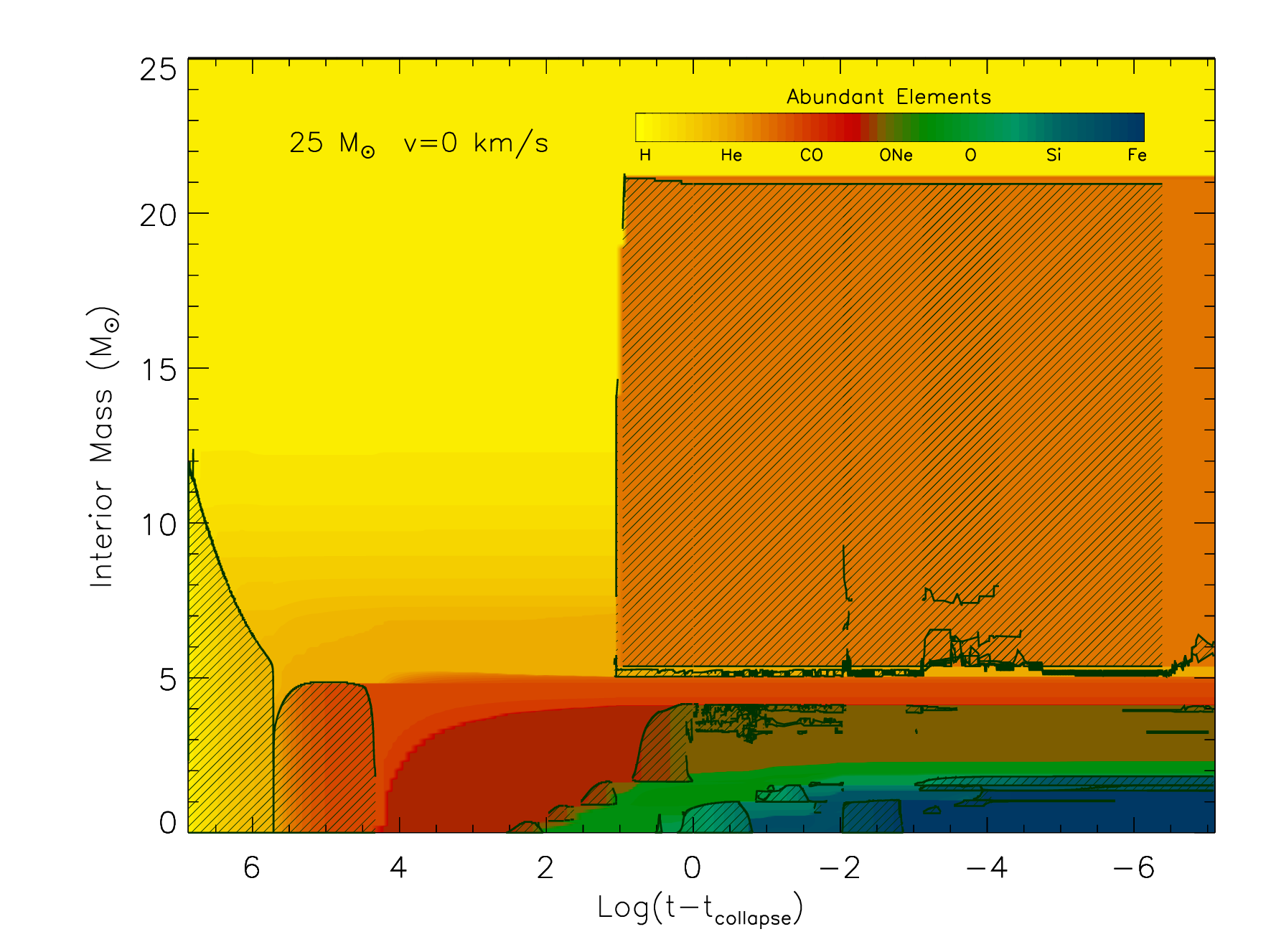}
\caption{Presupernova chemical and convective history of a $\rm 25÷M_\odot$. The shaded
regions correspond to the convective zones. The color gradient refers to the chemical composition as
reported in the colorbar}.\label{conv2}
\end{figure}

\begin{figure}
\epsscale{.80}
\plotone{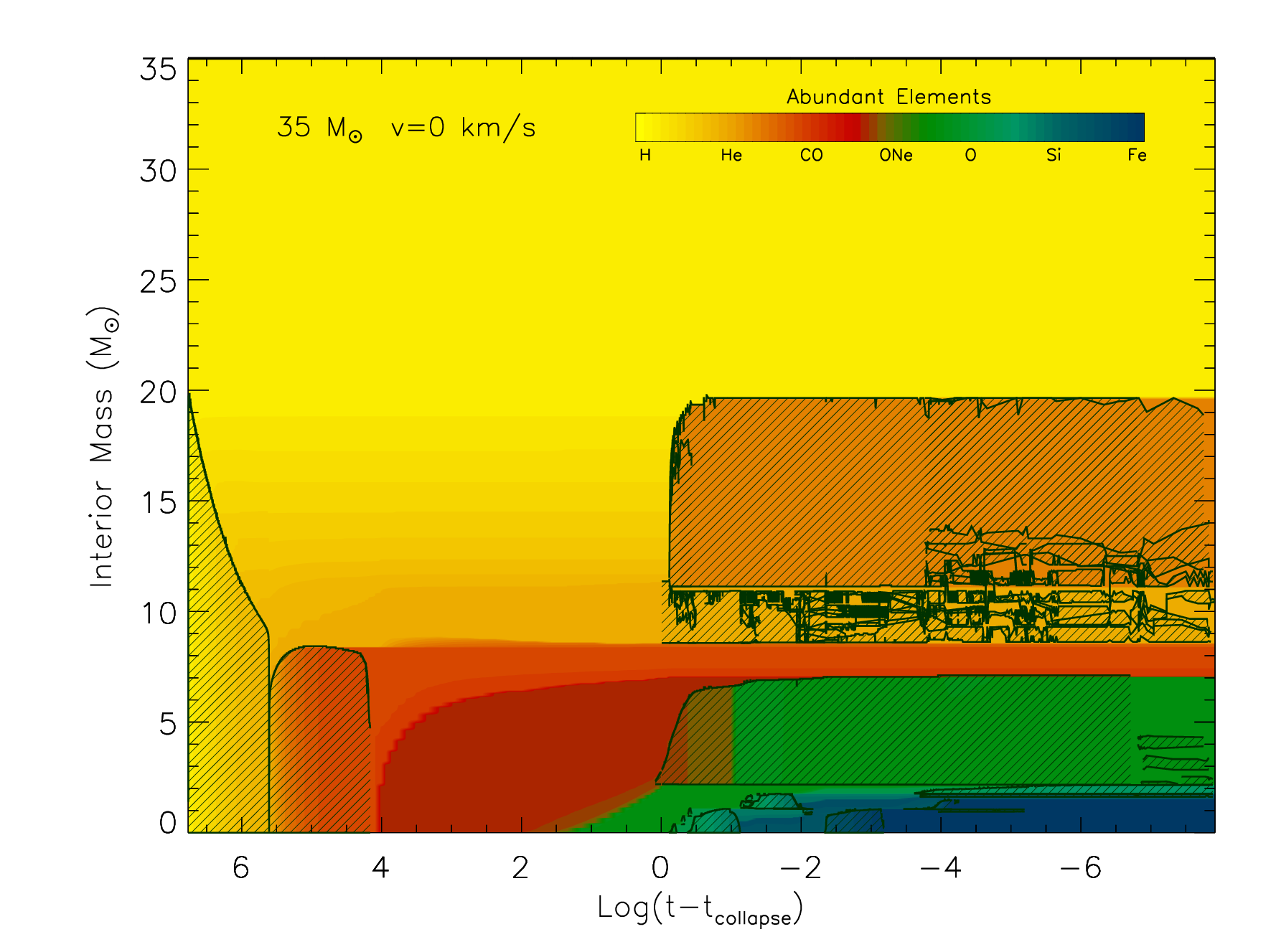}
\caption{Presupernova chemical and convective history of a $\rm 35÷M_\odot$. The shaded
regions correspond to the convective zones. The color gradient refers to the chemical composition as
reported in the colorbar}.\label{conv3}
\end{figure}

\begin{figure}
\epsscale{.80}
\plotone{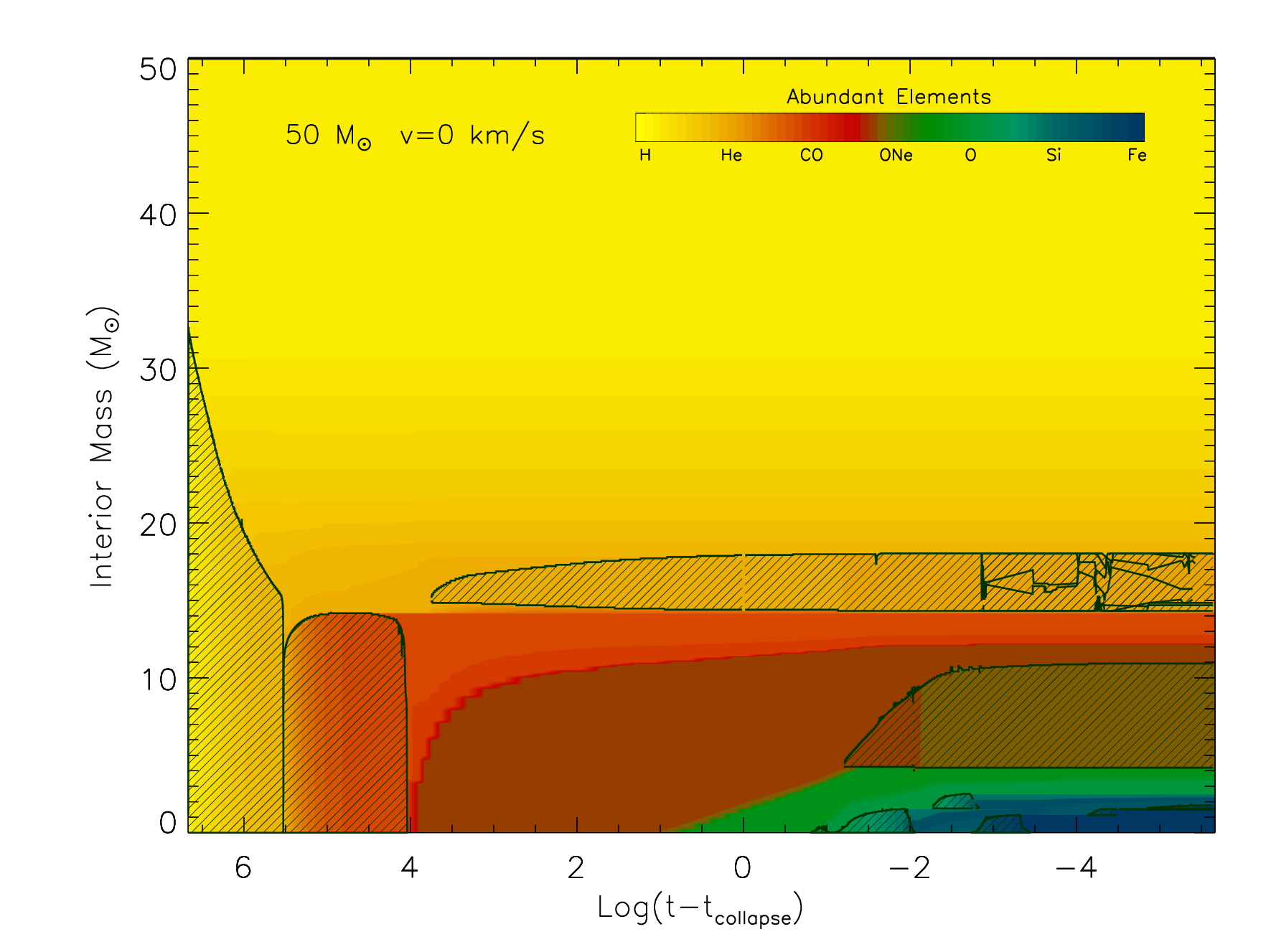}
\caption{Presupernova chemical and convective history of a $\rm 50÷M_\odot$. The shaded
regions correspond to the convective zones. The color gradient refers to the chemical composition as
reported in the colorbar}.\label{conv4}
\end{figure}

\begin{figure}
\epsscale{1.0}
\plotone{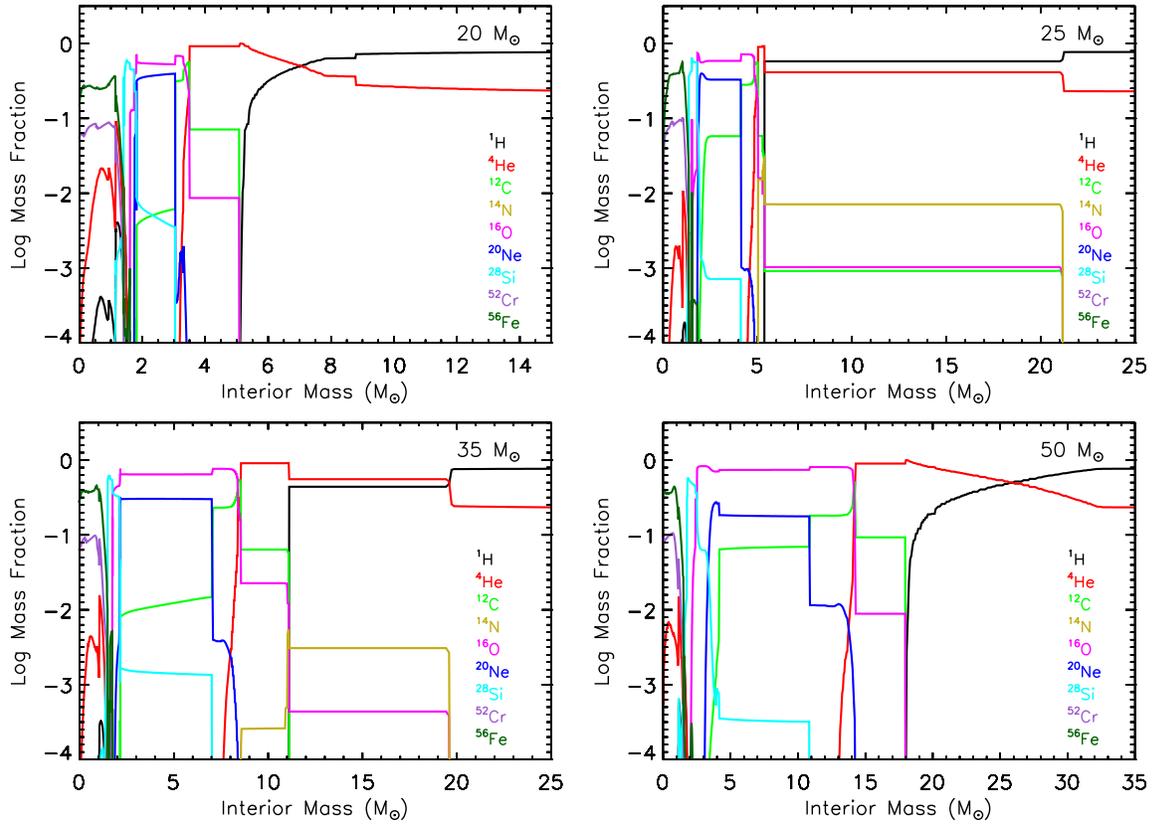}
\caption{Presupernova chemical structure of the 4 selected models presented in Figures \ref{conv1}, \ref{conv2}, \ref{conv3} and \ref{conv4}}.\label{major}
\end{figure}

Figures \ref{conv1}, \ref{conv2}, \ref{conv3} and \ref{conv4} show the presupernova chemical and convective histories of 4 representative cases ($\rm 20~M_\odot$, $\rm 25~M_\odot$, $\rm 35~M_\odot$ and $\rm 50~M_\odot$) selected among the full set of models. These figures clearly show that a very extended He convective shell develops in the mass range $\rm 25-35~M_\odot$. Such an extended convective shell is triggered by the proton ingestion described above and contains a substantial amount of primary N due to the C-rich environment in which the protons are ingested. The complex interplay, timing and overlap of the various convective zones and burning shells shapes the final presupernova structure given in Figure \ref{major} for the 4 previously selected models. Note the very extended He convective shell present in the models with mass $\rm 25~M_\odot$ and $\rm 35~M_\odot$ where both the H, ingested from the overlying zones, and the $\rm ^{14}N$, locally produced by the CNO cycle, are quite abundant. It is worth noting the reduction of the He core due to the proton ingestion which occurs during core C burning for the $\rm 25~M_\odot$ and the $\rm 30~M_\odot$ models. In these models the protons are mixed down to the base of the He convective shell hence the He core almost coincides with the CO core. In the $\rm 35~M_\odot$ model, the proton ingestion occurs during core Ne burning and in this case the He core remains well detached from the CO core. It is worth noting that
these results are consistent with those found in \citet{cl04}, although in this case
the primary $\rm ^{14}N$ production is increased by a factor of $\sim 2$ on average.
The origin of this increased production is probably due to the coupling between nuclear burning and convective mixing performed in these new calculations.

The basic evolutionary properties of all the stellar models during the presupernova evolution are reported in Table \ref{presnprop}.
More specifically, for each model, the following quantities are reported for each nuclear burning stage (column 1):
the nuclear burning lifetime in year (column 2); 
the average (photon) luminosity in $\rm erg~s^{-1}$ (column 3);
the average central temperature in Kelvin degrees (column 4);
the average central density in $\rm g~cm^{-3}$ (column 5);
the maximum mass of the convective core in $\rm M_\odot$ (column 6);
the final He core mass in $\rm M_\odot$ (column 7);
the final CO core mass in $\rm M_\odot$ (column 8);
the logarithm of the final effective temperature in Kelvin degrees (column 9);
the logarithm of the final luminosity in $\rm L_\odot$ (column 10).

\section{Explosive nucleosynthesis and calibration of the mass cut}
As in our previous computation of the explosion of a massive star, we induce the formation of a shock wave by imparting a velocity $v_0$ to a mass shell located at roughly 1~$\rm M_\odot$ from the center. As the shock propagates outward in mass it increases locally both the temperature and the density, triggering the explosive nucleosynthesis, but it also accelerates outward the shocked matter. However, not all layers will necessarily receive a kinetic energy large enough to overcome their binding energy. The more internal layers, in particular, being those with the largest binding energy, will be the first to fall back onto the remnant if the initial velocity is not high enough to allow the ejection of the whole mantle. Hence, for each given initial velocity $v_0$, a natural separation will occur between the part of the mantle that will fall back onto the remnant and the fraction really ejected in the interstellar medium. The mass cut ($\rm M_{cut}$) is defined as the mass coordinate which separates the final remnant from the ejected material. Once the shock wave reaches the surface of the exploding star, the further evolution of the mantle is characterized by a homologous expansion with a given total kinetic energy $\rm (E_{kin})$ which depends on the initial velocity $v_0$. Thus, in this framework, $\rm E_{kin}$ and $\rm M_{cut}$ are not independent quantities, but, on the contrary, the $\rm M_{cut}$ depends on the initial velocity $v_0$ and hence on the final $\rm E_{kin}$.

The lack of information about the initial properties of the shock wave (which should come from
a self consistent core collapse explosion simulation) as well as the very simple approximations (among which the spherical symmetry), make the $\rm E_{kin}$-$\rm M_{cut}$ relation very uncertain. Indeed, it may change, even significantly, depending on the various numerical/physical assumptions. For example, the adoption of a thermal bomb or a piston to start the explosion may lead to different masses of the remnant for the same final kinetic energy of the ejecta. A proper computation of the $\rm M_{cut}$ would require a much more sophisticated multidimensional hydrodynamical simulation of the core collapse explosion extending at late times that, at present, is not available.

Unfortunately, the yields of the nuclear species synthesized in the deepest part of the mantle depend dramatically on the mass cut for obvious reasons. In our previous papers we made the choice to either fix the $\rm M_{cut}$ in order to eject a given amount of $\rm ^{56}Ni$ \citep{cl02} or we provided tables of yields for different final kinetic energies of the ejecta and hence different values of the mass cut \citep{lc03,cl04}. In this paper, on the contrary, we adopt a different strategy taking advantage of the choice we made to fit the average star with the ejecta of a single supernova.
In particular, since for each single supernova there is a one to one relation between the mass cut (or equivalently the final kinetic energy) and the predicted $\rm [X/Fe]_{\rm model}$, we chose the mass cut so as to minimize the differences between $\rm [X/Fe]_{\rm model}$ and $\rm [X/Fe]_{\rm avg}$. We define the best fit to the observations in this way.

Figure \ref{fitavg} shows the best fit to the observed abundance pattern of the average star \citep{cayreletal04,spiteetal05} obtained with the present set of models following the above mentioned procedure.

We report in table \ref{tabyields}, for each exploded model, the iron core mass at the presupernova stage in solar masses ($\rm M_{Fe}$),
the explosion energy (in foe, 1 foe=$10^{51}$ erg), the related mass cut in solar masses ($\rm M_{cut}$) derived from the best fit procedure
and the ejected masses of each isotope in solar masses. Let us remind that the evolution of the chemical composition during the explosion
is followed up to a time of $\rm t=2.5\cdot 10^{4}~s$. Hence only the unstable isotopes with a very short half-life have time to decay in this time interval. Let us eventually mention that, although the yields provided in Table \ref{tabyields} have been obtained for a specific choice of the mass cut, the full set of cumulative isotopic yields as a function of the mass cut down to the Fe core mass are freely available at the web site http://www.iasf-roma.inaf.it./orfeo/public{\_}html. 

\begin{figure}
\epsscale{1.0}
\plottwo{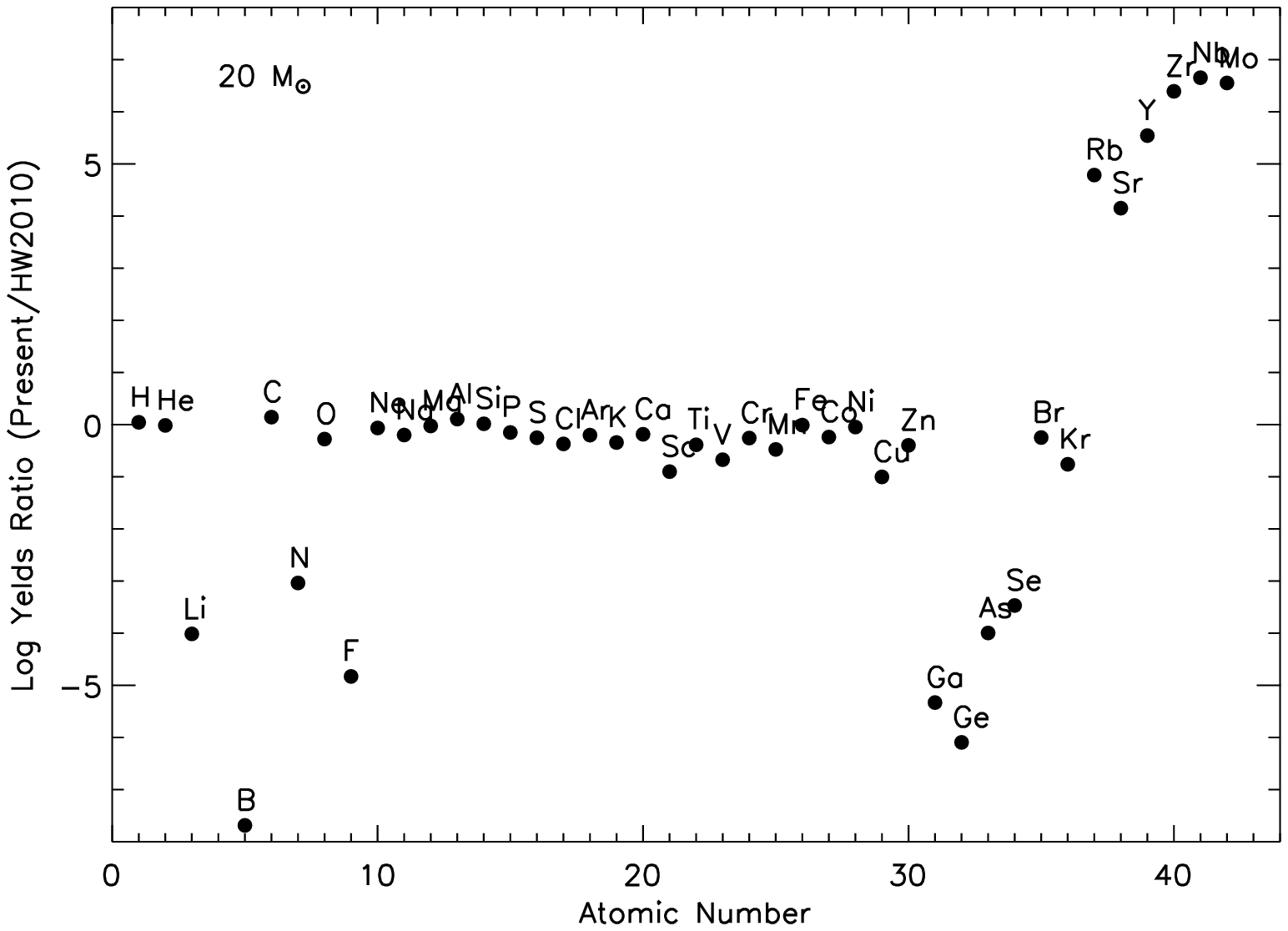}{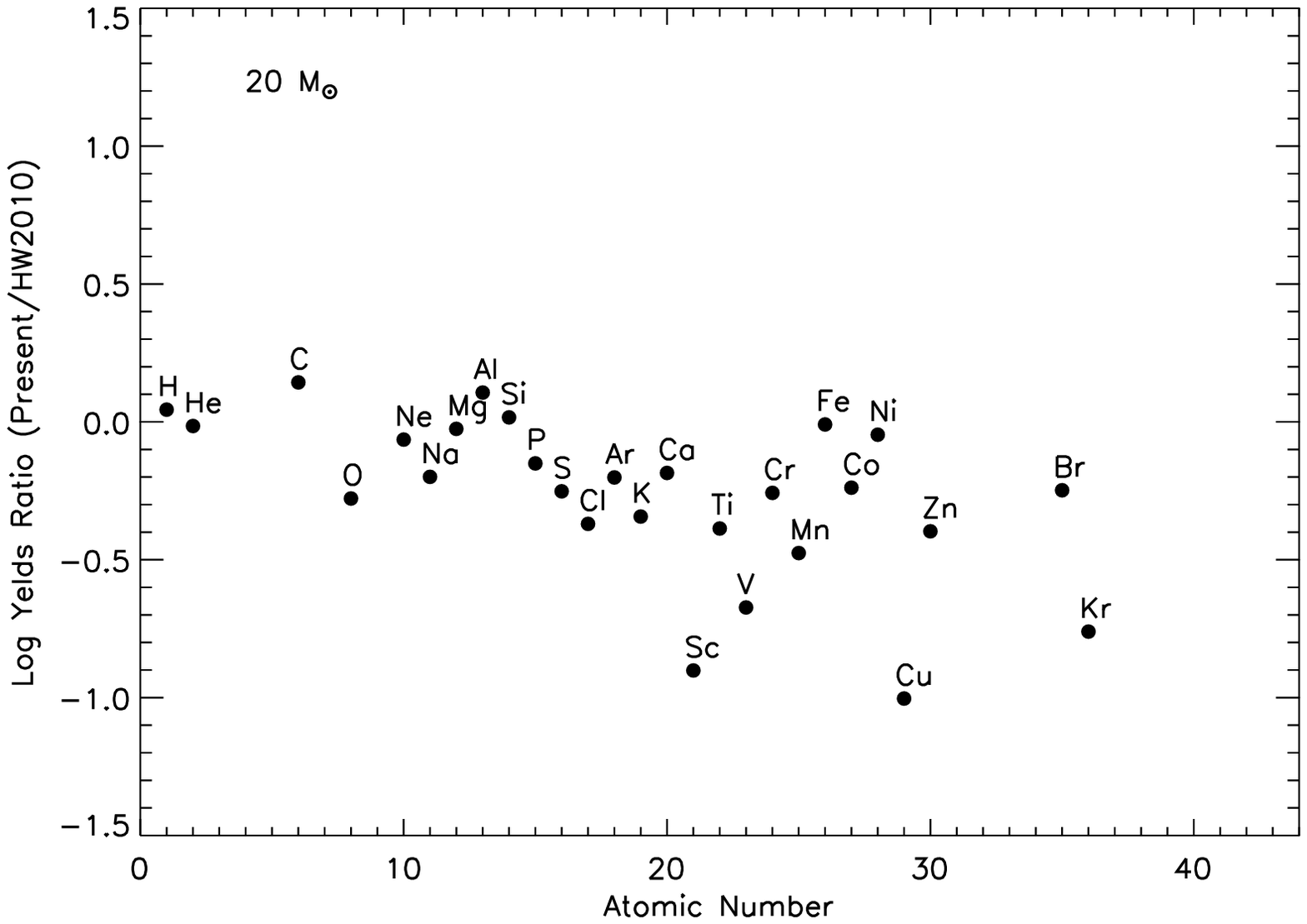}
\caption{Comparison between the yields produced by our 20 $\rm M_\odot$ and the analogous one provided by \cite{hw10}}.\label{conf_hw10}
\end{figure}

Our yields are certainly not the only ones on the "market" and hence it would be of extreme interest to try to understand where the differences among different sets of models come from. In the past we systematically compared our results with those obtained by other groups \citep{cls98,lsc99}. Unfortunately such an exercise did not allow us to understand the source of the differences. The reason is that there are so many differences in both the physical assumptions and the numerical techniques that a meaningful comparison would require a huge amount of work. Nonetheless, just to give a "flavour" of the differences that may be met by adopting one author or another, we show a comparison between the yields produced by a metal free 20 $\rm M_\odot$ and an analogous model presented by \cite{hw10} (hereinafter HW10). The dynamics of the explosion in our model was tuned so as to eject the same amount of $\rm ^{56}Ni$ ejected by the HW10 model. Figure \ref{conf_hw10} shows such a comparison: panel a) gives a global picture of the differences while panel b) focuses on a more expanded scale. It is evident that the HW10 model produces a larger amount of s-weak nuclei up to Kr while our model overproduces (with respect to HW10) nuclei between Kr and Mo. Also O, N and F appear overproduced (with respect to us) by HW10 while we produce more C. The block of elements between P and Fe are produced (on the average) in larger amount by the HW10 model. Without wanting to have the presumption to have identified the reason for the differences, let us simply suggest that a lower C abundance at the end of the central He burning phase in the HW10 model could explain at least part of the differences. It goes without saying that such an occurrence would directly lower the C yield and raise the O one. A lower C abundance in the CO core implies a faster advancing of the C shell and, in turn, a more robust contraction of the CO core itself \citep{lc06}. Since the He convective shell lies just above the CO core, a stronger contraction of the core itself triggers a stronger efficiency of the convective shell and hence a more generous penetration in the H rich mantle. Presumably also the timescale over which such an ingestion occurs shortens. The larger amount of protons ingested in the convective shell naturally leads to a larger production of N and F and to a stronger neutron flux. The consequence of such a strong n-flux, on a shorter timescale, is the preferential synthesis  of the lighter s-weak component (up to Kr) with respect to the more massive one (i.e. up to the N=50 neutron shell closure).

\section{Discussion and summary}
\begin{figure}
\epsscale{1.0}
\plotone{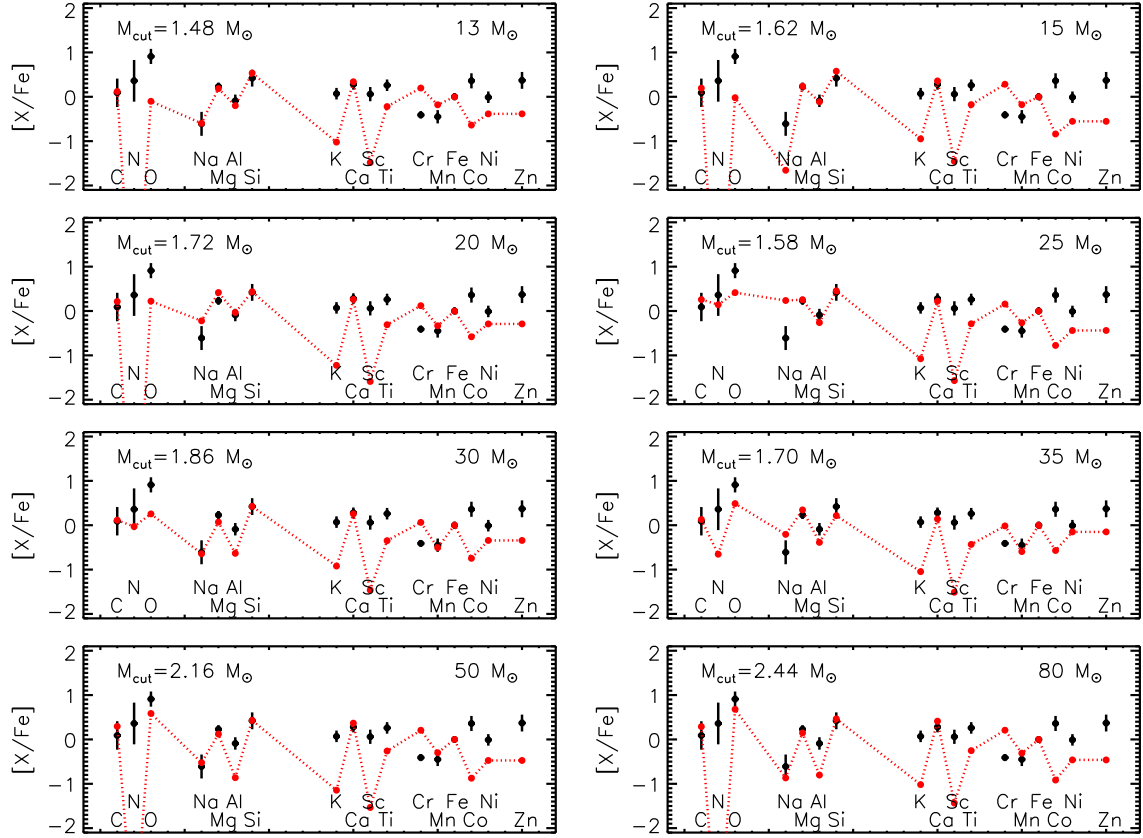}
\caption{Best fit to the observed abundances of the average star \citep{cayreletal04} obtained with the present set of models
following the procedure described in the text}.\label{fitavg}
\end{figure}

In this section we want to discuss in some detail the comparison between the theoretical yields
and the observed abundance pattern of the average star shown in Figure \ref{fitavg}.

First of all, let us discuss separately the element from C to Ca and from Sc to Zn.
The relative abundances of C, Mg, Si and Ca are very well reproduced by all the models. 
Since these elements are made by nuclear burning occurring either in the hydrostatic (C, Mg) and
in the explosive conditions (Si, Ca), this result is a good check of both the presupernova evolution
as well as of the explosive nucleosynthesis. The relative abundances of Mg, Al and Si is in general
well reproduced by the lower mass models ($\rm M<30~M_\odot$); the higher mass ones show a more pronounced odd-even
effect and hence they tend to underestimate the [Al/Fe] ratio compared to the observations.
The [Na/Fe] is in general well reproduced with the exception of the models
which undergo an efficient proton ingestion in the He convective shell ($\rm 20-25~M_\odot$) because they produce a 
substantial amount of primary Na and hence tend to overestimate the observed [Na/Fe].
The best fit of the relative abundances of C, Na, Mg, Al and Ca is obtained for the $\rm 13~M_\odot$.
The origin of the high odd-even effect, which is the cause of the very low Al abundance compared to Mg and Si
in the more massive models, is not clear at the moment.
The [N/Fe] is severely underestimated by all the models except those in which the primary nitrogen
production is activated by the ingestion of protons in the He convective shell. The best fit to the observed
[N/Fe] is obtained for models in the mass range $\rm 20-25~M_\odot$.
The [O/Fe] ratio increases with increasing the mass since it is mainly produced by the core He burning and hence it
scales with the size of the He core mass. The large observed value of [O/Fe] is reproduced only by the
more massive model, i.e., the $\rm 80~M_\odot$. [K/Fe] is always severely underproduced by all the models.

Overall, the observed element abundance ratios of [C,Mg,Si,Ca/Fe] is very
well reproduced by all the models; the almost solar abundance of [Al/Fe] points toward a low mass
progenitors while the large [O/Fe] ratio is compatible with high mass models. The large observed [N/Fe], which is probably
of primary origin, is obtained only for intermediate mass models which, however, overestimate substantially the
observed [Na/Fe]. Therefore, there is no single model, nor a generation of massive stars, 
that can reproduce simultaneously ALL the C to Ca abundance pattern observed in the average star.

The comparison between the observed and the predicted abundances of the elements from Sc to Zn
(Sc, Ti, Cr, Mn, Co, Ni and Zn) confirms some well known problems \citep{cl02,lc02,un02,un03,un05}:
the abundances of Sc, Ti, Co, Ni and Zn relative to Fe are significantly underestimated while [Cr/Fe] is overproduced compared to the observations.
In this set of models the [Mn/Fe] ratio is always very well reproduced. Since Mn and Cr are produced in the same
zone by the incomplete explosive Si burning, the fact that only one of the two elements can be well
reproduced by the models constitute an additional problem. As already mentioned above, only
non standard models, in which parametrized mixing and fall back mechanism, very energetic and/or aspherical explosions
are assumed, can reproduce better the heavy elements abundance pattern. 

In summary, we computed a set of zero metallicity massive star models and their related explosive nucleosynthesis
in the framework of the induced explosion, with typical energies of the order of $\rm 10^{51}~erg$ and without
adopting any artificial mixing and fallback during the explosion. We calibrate the kinetic energy (and consequently the mass cut) of each model in order to minimize the differences between the predicted and the observed abundances in extremely metal poor stars.
We confirm the well known result that there is no single model, nor a generation of zero metallicity massive stars, 
that can reproduce simultaneously ALL the element abundance pattern observed in the average star. The only way in which at present it is possible to improve the fit to the observed [X/Fe] in the average star is to assume properly tuned unconventional explosions
(large energies, mixing and fallback, aspherical explosions, $\rm Y_e > 0.5$, etc.) \citep[see, e.g.][]{un05,tun07} and properly tuned IMFs \citep{hw10}. 
Also in this case, however, a fully satisfactory match to the observational data is not completely achieved.
The role of rotation could certainly play an important role, especially in the primary production of N,
but a comprehensive study involving rotating models in which the abundance pattern of ALL the observed elements are considered
is not available yet \citep{meynetetal10}. We plan to address such a problem filling this gap, quite shortly.

\clearpage

\clearpage

\clearpage

\clearpage

\clearpage

\clearpage

\clearpage

\clearpage

\clearpage

\begin{deluxetable}{lccccccccc}
\tabletypesize{\scriptsize}
\tablecolumns{10}
\tablewidth{0pc}
\tablecaption{Key evolutionary properties.\label{presnprop}}
\tablehead{
\colhead{Phase} & \colhead{time}   & \colhead{$\rm <L_\gamma>$}    & \colhead{$\rm <T_c>$} &
\colhead{$\rm <\rho_c>$}    & \colhead{$\rm M_{cc}$}   & \colhead{$\rm M_{He}$}    & \colhead{$\rm M_{CO}$} & \colhead{$\rm T_{eff}$}    & \colhead{$\rm Log~L/L_\odot$}}
\startdata
\cutinhead{$\rm 13~M_\odot$}
  H    & $1.34\cdot 10^{ 7}$ &  $7.98\cdot 10^{37}$ &  $9.78\cdot 10^{7}$ &  $1.79\cdot 10^{2}$ & $3.26$        &  $2.06$ &  $0.00$ &   $4.65$ &  $4.49$   \\
  He   & $1.08\cdot 10^{ 6}$ &  $1.34\cdot 10^{38}$ &  $1.87\cdot 10^{8}$ &  $1.48\cdot 10^{3}$ & $1.53$        &  $2.86$ &  $1.52$ &   $4.62$ &  $4.56$   \\
  C    & $2.48\cdot 10^{ 4}$ &  $1.32\cdot 10^{38}$ &  $6.10\cdot 10^{8}$ &  $5.26\cdot 10^{5}$ & $0.65$        &  $3.02$ &  $1.77$ &   $4.32$ &  $4.52$   \\
  Ne   & $2.56\cdot 10^{ 1}$ &  $1.26\cdot 10^{38}$ &  $1.04\cdot 10^{9}$ &  $1.75\cdot 10^{7}$ & $0.29$        &  $3.02$ &  $1.80$ &   $4.32$ &  $4.52$   \\
  O    & $1.03\cdot 10^{ 1}$ &  $1.26\cdot 10^{38}$ &  $1.72\cdot 10^{9}$ &  $1.50\cdot 10^{7}$ & $0.88$        &  $3.02$ &  $1.79$ &   $4.32$ &  $4.52$   \\
  Si   & $6.88\cdot 10^{-1}$ &  $1.26\cdot 10^{38}$ &  $2.02\cdot 10^{9}$ &  $9.57\cdot 10^{7}$ & $0.00$        &  $3.02$ &  $1.79$ &   $4.32$ &  $4.52$   \\
 PSN   & $7.25\cdot 10^{-2}$ &  $1.26\cdot 10^{38}$ &  $3.18\cdot 10^{9}$ &  $1.38\cdot 10^{8}$ & \nodata\phn   &  $3.02$ &  $1.75$ &   $4.32$ &  $4.52$   \\
\cutinhead{$\rm 15~M_\odot$}
  H    & $1.12\cdot 10^{ 7}$ &  $1.16\cdot 10^{38}$ &  $1.00\cdot 10^{8}$ &  $1.57\cdot 10^{2}$ & $4.54$         &  $2.57$ &  $0.00$ &   $4.67$ &  $4.66$   \\
\enddata
\end{deluxetable}

\clearpage

\begin{deluxetable}{lcccccccc}
\tabletypesize{\scriptsize}
\tablecolumns{9}
\tablewidth{0pc}
\tablecaption{Explosion properties and ejected masses (in $\rm M_\odot$ and $\rm 2.5\cdot 10^4~s$ after the explosion) for all the computed models \label{tabyields}}
\tablehead{
\colhead{} & \colhead{$\rm 	13$}   & \colhead{$\rm 15$}  & \colhead{$\rm 20$}    & \colhead{$\rm 25$} &
\colhead{$\rm 30$} & \colhead{$\rm 35$}   & \colhead{$\rm 50$}  & \colhead{$\rm 80$}}
\startdata
$\rm M_{Fe} $   &     1.34            &    1.40               &     1.40              &     1.36              &    1.46                 &    1.50                &     1.56               &     2.08  \\
$\rm E_{kin}$  &     0.50             &    0.74               &     0.97              &     1.10              &    1.58                 &    1.28                &     2.59               &     5.22  \\
$\rm M_{cut}$  &     1.48             &    1.62               &     1.72              &     1.58              &    1.86                 &    1.70                &     2.16               &     2.44  \\
$\rm ^{ 1}H $  & $7.05\cdot 10^{+00}$ &  $8.01\cdot 10^{+00}$ & $1.01\cdot 10^{+01}$  &  $2.74\cdot 10^{+00}$ &   $3.58\cdot 10^{+00}$  &  $1.16\cdot 10^{+01}$  &  $2.00\cdot 10^{+01}$  &  $2.87\cdot 10^{+01}$ \\
$\rm ^{ 2}H $  & $9.03\cdot 10^{-17}$ &  $8.29\cdot 10^{-17}$ & $9.88\cdot 10^{-17}$  &  $1.90\cdot 10^{-16}$ &   $3.19\cdot 10^{-16}$  &  $4.67\cdot 10^{-16}$  &  $1.08\cdot 10^{-15}$  &  $1.47\cdot 10^{-14}$ \\
$\rm ^{ 3}H $  & $1.33\cdot 10^{-29}$ &  $6.99\cdot 10^{-29}$ & $8.66\cdot 10^{-28}$  &  $8.70\cdot 10^{-27}$ &   $2.54\cdot 10^{-25}$  &  $2.50\cdot 10^{-25}$  &  $1.04\cdot 10^{-24}$  &  $6.68\cdot 10^{-22}$ \\
$\rm ^{ 3}He$  & $3.12\cdot 10^{-05}$ &  $3.04\cdot 10^{-05}$ & $2.93\cdot 10^{-05}$  &  $2.81\cdot 10^{-05}$ &   $2.79\cdot 10^{-05}$  &  $2.78\cdot 10^{-05}$  &  $2.81\cdot 10^{-05}$  &  $3.00\cdot 10^{-05}$ \\
$\rm ^{ 4}He$  & $4.11\cdot 10^{+00}$ &  $4.71\cdot 10^{+00}$ & $6.29\cdot 10^{+00}$  &  $7.79\cdot 10^{+00}$ &   $9.31\cdot 10^{+00}$  &  $1.07\cdot 10^{+01}$  &  $1.54\cdot 10^{+01}$  &  $2.43\cdot 10^{+01}$ \\
\enddata
\end{deluxetable}

\end{document}